\documentclass[aps,prb]{revtex4}
\usepackage{amsmath,amssymb}
\usepackage{graphicx}
\usepackage{dcolumn}
\usepackage{bm}
\usepackage{color}

\newcommand{\mc}{\mathcal}

\newcommand{\tet}{\texttt}
\newcommand{\pr}{\partial}

\begin{document}

\title{
Coherent-scatterer enhancement and Klein-tunneling suppression by potential barriers in gapped graphene with chirality-time-reversal symmetry}

\author{Farhana Anwar$^{1}\footnote{Email: notfarhana@gmail.com, fanwar@lbl.gov}$, Andrii Iurov$^{2}\footnote{Email: aiurov@mec.cuny.edu, theorist.physics@gmail.com}$, Danhong Huang$^{3}$, Godfrey Gumbs$^{4,5}$, and Ashwani Sharma$^{3,6,7}$}
\affiliation{
$^{1}$Lawrence Berkeley National Laboratory, 1 Cyclotron Rd, Berkeley, CA 94720\\
$^{2}$Department of Physics and Computer Science, Medgar Evers College of the City University of New York, Brooklyn, NY 11225, USA\\
$^{3}$Air Force Research Laboratory, Space Vehicles Directorate, Kirtland Air Force Base, NM 87117, USA\\
$^{4}$Department of Physics and Astronomy, Hunter College of the City University of New York, 695 Park Avenue, New York, NY 10065, USA\\
$^{5}$Donostia International Physics Center (DIPC), P de Manuel Lardizabal, 4, 20018 San Sebastian, Basque Country, Spain\\
$^{6}$Center for High Technology Materials, University of New Mexico, 1313 Goddard SE, Albuquerque, NM 87106, USA\\
$^{7}$Department of Electrical and Computer Engineering, University of New Mexico, Albuquerque, NM 87106, USA
}
\date{\today}

\begin{abstract}
We have utilized the finite-difference method to investigate electron tunneling in gapped graphene across various electrostatic-potential barriers.  Specifically, we analyze Gaussian and triangular envelope functions to  compare with a square potential barrier. The transmission coefficient was calculated numerically for each case and applied to corresponding tunneling conductance. It is well known that Klein tunneling in graphene will be greatly reduced in a gapped graphene. Our results further demonstrated that such a  decrease of transmission can be significantly enhanced for spatially-modulated potential barriers. Additionally, we investigated the effect from a bias field applied to those barrier profiles, from which we showed that it enables the control of electron flow under  normal incidence. The suppression of Klein tunneling was found to be more severe for a non-square barrier and exhibits a strong dependence on bias-field polarity for all kinds of barriers. The roles played by a dilute distribution of point impurities on electron transmission and conductance were analyzed with a sharp peak appearing in the electron conductance as an impurity atom is placed in the middle of a square barrier.  However, for narrow triangular and Gaussian barriers,  the conductance peaks become substantially broadened,  associated with an enhancement of the tunneling conductance.\end{abstract}

\maketitle

\section{Introduction}
\label{s1}

The  unusual properties of Dirac quasiparticles have become one of the most popular topics in fundamental research as well as a promising source for new technological applications \,\cite{geimrise,neto2009electronic,geim2009graphene}. Being a zero-bandgap semiconductor and possessing a specific chiral wave function simultaneously will result in full transparency for any potential barrier for normally-incident electrons.\,\cite{beenakker2008colloquium,kats1} A massless Dirac fermion which is scattered by an electrostatic potential is able to tunnel with certainty for normal incidence, as known Klein paradox, regardless of the potential-barrier height or width.\,\cite{klein1929reflexion,calogeracos1999history} Graphene is also known for its specific magnetic-field properties,\,\cite{mag0,butter,gus1,mag02} as well as its strong decrease of magneto-resistance\,\cite{dh19, mag2} through $p$-$n$ junctions.\,\cite{mag1}   While the metallic band structure and high mobility are obvious advantages of graphene, most applications in electronics require a small and tunable bandgap\,\cite{pereira2008supercritical} in order to keep its charged carriers confined within a finite area of an electronic device.\,\cite{berger2006electronic} This feature could be achieved by opening a finite bandgap\,\cite{ni2008uniaxial} which directly leads to a suppression of the Klein tunneling in graphene.\,\cite{iurov2011anomalous,kindermann2012zero,zhou2007substrate,our2}

\medskip
\par

There are various  efficient ways for creating a sizable ($\sim 100\,$meV and above) and tunable bandgap in graphene. Most of them are generated  byadding a dielectric Si-based substrate\,\cite{ni2008uniaxial} or a substrate with broken inversion symmetry between sublattices.\,\cite{zhou2007substrate} It was just recently discovered that a tunable gap can be achieved after a Dirac-cone hexagonal two-dimensional (2D) lattice has been irradiated with a circularly-polarized\,\cite{ourpeculiar,oka2009photovoltaic,liu2011visualizing} and off-resonance field\,\cite{manyb2020,pastrana2012advanced,usaj2014irradiated,ourfloqjap17,k1,k2,k3,k4,iurov2017exchange,dey2018photoinduced}. In addition, strain engineering, finite-width graphene nanoribbon (GNR) systems,\,\cite{brey2006electronic}, structural and topological defects or insertion of impurity atoms can also produce a bandgap in graphene.\,\cite{pereira2008supercritical} Physically, the presence and size of an induced bandgap in graphene and other hexagon lattices directly affect tunneling and transport properties\,\cite{kisrep,our2020}, and play a crucial role in graphene-based electronics, such as field-effect transistors since their on/off current ratios\,\cite{low2009electronic} can be tuned through tunneling control. Also, the optical response of gapped graphene   acquire attractive features in opto-electronics and optical spectroscopy.\,\cite{pedersen2009optical} Particularly, these responses have been proved to be sensitive to localized and trapped states within the bandgap of systems considered.\,\cite{pereira2006disorder,swj14,castro2008localized} The effect due to mesoscopic fluctuations appearing in the conductance of a gapped-graphene strip  arising from a random electrostatic potential landscape is believed to be important but has not been investigated thoroughly as for gapless graphene.\,\cite{rycerz2007anomalously} The same goes for the effect of an electrostatic field applied across a potential barrier.\,\cite{jang2013graphene}
\medskip 
\par

Although a considerable amount of research has been published on graphene band-transport characteristics, investigations on tunneling  across a smoothly-varying potential barrier in a gapped Dirac system have been much less dealt with. In fact, tunneling transport through a steep-slope potential profile, involving either single or double square barrier, has been investigated and reported in Refs.\, [\onlinecite{dahal2017effect,azarova2014transport,navarro2018bandgap}]. However, analytical solutions for a finite-slope barrier are still inaccessible. On the other hand, electron transmission in graphene and other newly discovered Dirac materials was also computed based on Wentzel-Kramers-Brillouin (WKB) semiclassical theory.\,\cite{sonin,wkb} Furthermore, it has been shown within WKB theoru  that if  electron-hole transition is considered inside a non-square potential barrier, tunneling transmission will increase with the slope of a potential profile. Technically, a smooth finite-slope potential barrier becomes more realistic since it matches better with the experimentally accessible situation\,\cite{stander2009evidence}. Importantly, such a case also presents several intriguing phenomena and properties, e.g., Klein collimator\,\cite{libisch2017veselago} as well as a possibility for building up a collimated interferometer or reflector\,\cite{young2009quantum,wang2019graphene}. Besides, these smooth-barrier profiles also display unusual tunneling features upon applying an electric or a magnetic field.\,\cite{mouhafid2013transport,shytov2008klein,young2009quantum,anwar2017nanoscale,anwar2020tunneling}

\medskip
\par

Compared with tunneling transport of incident electrons through a potential barrier in graphene, impurity scattering\,\cite{titov2007impurity,ando1984field,rycerz2007anomalously,anwar2020interplay} in gapped graphene turns out to be another important research topic that has not been adequately studied. Disorder embedded within an electrostatic-potential barrier in a gapped graphene system \cite{huang1999effects}, which exhibits either disorder-assisted or disorder-impeded tunneling, is even less known. Specifically, the spatial position, strength and polarity of an embedded scatterer and its effect on tunneling conductance of electrons for various electrostatic potential barriers\,\cite{rycerz2007anomalously,titov2007impurity} have become the most important aspect because they decide whether the conductance of the system is enhanced or reduced by different impurity configurations. As a result, it is of paramount importance to study the effect of impurity scattering in gapped graphene and reveal the condition for suppressed back-scattering of incident electrons by appropriately distributing scatterers within a barrier region, and this makes a crucial part of our current work.

\medskip
\par

The key issue addressed in this paper is quantifying electron transmission and conductance across non-square potential barriers with a finite slope in gapped graphene subject to a DC electric field by using a discretized Dirac equation\,\cite{hernandez2012finite,tworzydlo2008finite} based on a finite-difference method\,\cite{anwar2020interplay,huang2020tunneling}. Our numerical method presented here has proven to be crucial for studying a gapped-graphene system with a smooth disorder potential. Based on our established numerical procedure, we hsvr accurately  calculate the effect of a disorder potential on electron tunneling and conductance. The remainder of the paper is organized as follows. We first validate our discretized model by calculating the transmission coefficient of gapped graphene across a potential barrier, and then compare the obtained numerical results with previous analytical solutions\,\cite{setare2010klein} in some limiting cases. By employing this numerical method, we further explore quantitatively how the barrier-potential profile, applied bias with different polarities and barrier-embedded disorders affect the conductance as a function of incident-electron kinetic energy for various graphene $p$-$n$ junction (GPNJ) within a gapped monolayer graphene.

\section{General Formalism}
\label{s2}

The low-energy states of gapped graphene can be described fully by a Dirac Hamiltonian with an additional gap term,\,\cite{setare2010klein, iurov2011anomalous, pavlo1} given by 

\begin{equation}
\label{generalHam}
\hat{\mc{H}} (\mbox{\boldmath$r$})=-i v_F\, \left[ \hat{\mbox{\boldmath$\Sigma$}} \cdot \mbox{\boldmath$\nabla$} \right]_{(x,y)}
+ V_B(x)\,\hat{\Sigma}_{0} + \Delta_G\,\hat{\Sigma}_z\ , 
\end{equation}
where $\hat{\Sigma}_{x,y,z}$ are three two-dimensional Pauli matrices, $\hat{\Sigma}_{0}$ is a $2 \times 2$ unit matrix, $v_F$ is the Fermi velocity, $\Delta_G$ is the gap parameter, and $V_B(x)$ represents
a spatially non-uniform barrier potential. For constant potential $V_B(x)=V_0$, the Dirac Hamiltonian in Eq.\,\eqref{generalHam} gives rise to a finite energy bandgap $E_G= 2\Delta_G$ between the valence and conduction bands which are symmetric with respect to the Dirac point. In addition, the energy dispersions are calculated as $\varepsilon_\gamma(k) = \gamma \sqrt{(\hbar v_F k)^2 + \Delta_G^2}$ and $\mbox{\boldmath$k$}=\{k_x,k_y\}$ is a two-dimensional wave vector of electrons. For varied $V_B(x)$, however, we have $\gamma=\text{sign}[\varepsilon_\gamma(k)-V_B(x)]=\pm 1$, corresponding to electron and hole states, respectively, where $\text{sign}(x)$ is a sign function. This implies that the carrier could go through an electron-hole (or inverse) transition inside the barrier region. 
\medskip 

In this paper, we  look forward to finding scattering-state solutions $\Phi_{\gamma}(\mbox{\boldmath$r$})$ for the Hamiltonian in Eq.\,\eqref{generalHam} in the form two-component (spinor) type of wave function, i.e.

\begin{equation}
\Phi_{\gamma}(\mbox{\boldmath$r$})= \tet{exp}(i k_y y)\,\Psi_{\gamma}(x)=\, \tet{exp}(i k_y y)\,
\left[
\begin{array}{c}
\phi_A (\gamma, x) \\
\phi_B (\gamma, x)
\end{array} 
\right] \ ,
\end{equation}
where unlike gapless graphene both components rely on the electron-hole index $\gamma$. 
\medskip 

As a special case, if $V_B(x)=V_0$ is a constant, the translational symmetry of the system is preserved in both $x$ and $y$ directions, and then the  
Hamiltonian in Eq.\,\eqref{generalHam} can be greatly simplified as

\begin{equation}
\label{gap01}
\hat{\mc{H}}^{(0)}_g(k \, \vert \, \theta_{\bf k}) = \left[
\begin{array}{cc}
V_0+\Delta_G & \hbar v_Fk_-\\
\\
\hbar v_Fk_+ & V_0-\Delta_G
\end{array}
\right]\ , 
\end{equation}
where $\mbox{\boldmath$k$}=(k_x^{(0)},k_y)$, $k_{\pm} = k_x^{(0)} \pm i k_y$, $\gamma=\text{sign}(\varepsilon_0(k)-V_0)$ within the barrier region, and $\varepsilon_0(k)=\hbar v_F\sqrt{(k_x^{(0)})^2+k_y^2}$ is the given energy of an incident electron. In this case, the scattering-state wave function associated with the Hamiltonian in Eq.\,\eqref{gap01} takes the explicit form\,\cite{pavlo1,iurov2011anomalous,ourjapc}

\begin{equation}
\Psi_{\gamma}^{(0)}(\mbox{\boldmath$r$}) =\frac{1}{\sqrt{2\gamma\,\delta\varepsilon_0(k)}} \left[
\begin{array}{c}
\sqrt{\vert \delta\varepsilon_0(k)+\Delta_G\vert}\\
\\
\gamma\sqrt{\vert\delta\varepsilon_0(k)-\Delta_G\vert}\,\tet{e}^{i\theta_{\bf k}}
\end{array}
\right]\tet{exp}(i k^{(0)}_x x+ik_y y)\ ,
\label{wfunc}
\end{equation}
where $\theta_{\bf k}=\tan^{-1}(k_y/k_x^{(0)})$, $\delta\varepsilon_0(k)\equiv\varepsilon_0(k)-V_0\geq\Delta_G$ for an electron with $\gamma=+1$ and  
$\delta\varepsilon_0(k)\leq-\Delta_G$ for a hole with $\gamma=-1$. Here, two components of the wave function spinor in Eq.\,\eqref{wfunc}
are not the same and their ratio further depends on the electron/hole index $\gamma = \pm 1$.  
\medskip 
\par

\begin{figure}	
\centering
\includegraphics[width=0.7\textwidth]{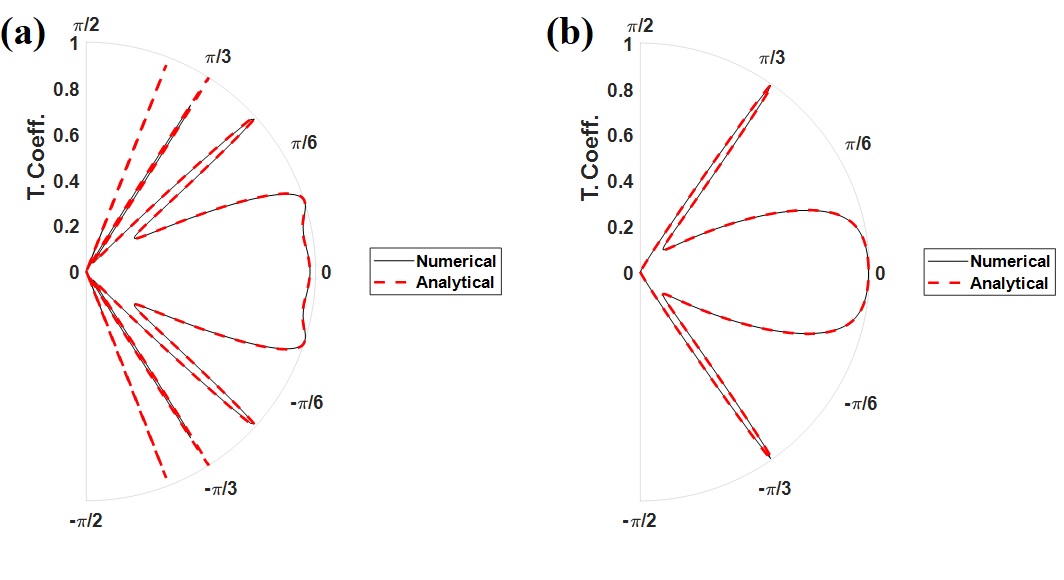}
\caption{(color online) Polar plots for the transmission coefficient $T(\varepsilon, \phi_{\bf k} \, \vert \, W_B)$ as a function of incidence angle $\phi_{\bf k}$ in graphene with a bandgap $\Delta_G=50\,$meV for a square potential barrier $V_s(x)=V_0\,\Theta(x)\,\Theta(W_B-x)$ with various barrier widths $W_B$. Here, both the analytical expressions (solid curves) and numerical results (dashed curves) obtained by using the FDA method are plotted together for comparisons. Panel $(a)$ displays calculated $T(\varepsilon, \phi_{\bf k} \, \vert \, W_B)$ for $W_B=110\,$nm, while panel $(b)$ shows $T(\varepsilon, \phi_{\bf k} \, \vert \, W_B)$ for $W_B=50\,$nm, where the barrier height is $V_0=285\,$meV, the bias field ${\cal E}_{0}=0$, and the incident-electron energy $\varepsilon=80\,$meV are assumed. 
}
\label{f1}
\end{figure}

Here, we concentrate on studying the electron transmission and conductance through a biased barrier of different geometries in gapped graphene. Therefore, we  limit our investigation only to experimentally-available symmetrical barrier profiles, e.g., triangular and Gaussian shapes. For this reason, we would like to compare our results first with the well-studied case having a square barrier

\begin{eqnarray}
\label{finalt0}
&& \frac{V_s(x)}{V_0} = \Theta(x) \, \Theta(W_B - x) \, ,
\end{eqnarray}
where $\Theta(x)$ is the Heaviside step function, meaning $V_s(x) = V_0$ if $0 < x < W_B$ and is equal to zero otherwise. As adopted in Figs.\,\ref{f1} and \ref{f02}, the selection of using a square potential barrier makes it simple to compare our numerical results for a general varying-slope barrier system with some known analytical cases and discern quantitatively the effect of a finite slope of a potential barrier.

\medskip 
\par

\begin{figure}
\centering
\includegraphics[width=0.7\textwidth]{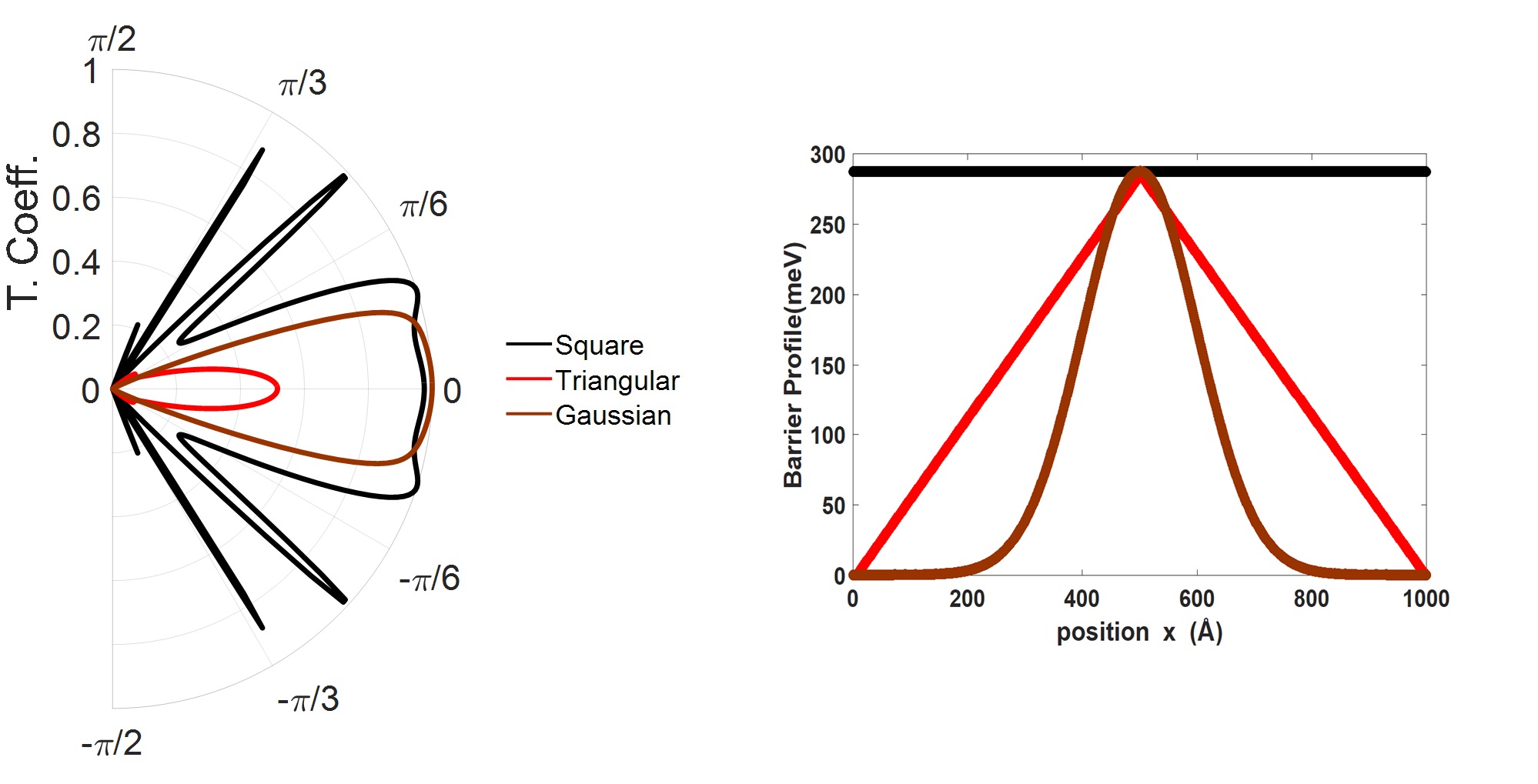}
\caption{(color online) (Left) comparison of the FDA calculated transmission coefficients $T(\varepsilon, \phi_{\bf k} )$ as a function of the angle of incidence $\phi_{\bf k}$ for three different unbiased barrier-potential profiles with $V_0=285\,$meV, ${\cal E}_{0}=0$, $\Delta_G=50\,$meV, and $\varepsilon=80\,$meV, including a square (black), a triangular (red) and Gaussian (brown) potential barrier. (Right) corresponding potential-barrier profiles $V_B(x)$ chosen for our computation results presented in the left panel.}
\label{f2}
\end{figure}

Next, for comparison we would introduce a triangular barrier $V_t(x)$ with both positive and negative slopes present on its two sides, yielding 

\begin{equation}
\label{finalt}
\frac{V_t(x)}{V_0} = \left\{ 
\begin{array}{ccc}
0           & \hskip0.1in \text{for} & \hskip0.1in x < a \ ,\\
(x-a)/(b-a) & \hskip0.1in \text{for} & \hskip0.1in a < x < b \ ,\\
(c-x)/(c-b) & \hskip0.1in \text{for} & \hskip0.1in b < x < c \ ,\\
0           & \hskip0.1in \text{for} & \hskip0.1in  x > c \ .\\
\end{array} 
\right.
\end{equation}
For our numerical computations, we  adopt a symmetric triangular barrier configuration which corresponds to $c-b = b-a$, as used in Fig.\,\ref{f2}.
\medskip

For comparison with experimentally-available setup, we also consider a (symmetric) Gaussian profile with the potential voltage profile varying according to 

\begin{eqnarray}
\label{finalt2}
&& \frac{V_g (x)}{V_0} = \pm\tet{exp}\left[-\left( \frac{x-x_0}{l} \right)^2 \right] \, ,
\end{eqnarray}
where $l$ is employed to define the effective width $\sqrt{\pi} l$ of our barrier, $x_0$ is its symmetric center where the highest possible potential
$V_0$ is reached, as employed in Fig.\,\ref{f2}. Depending on the $\pm$ sign in Eq.~\eqref{finalt2}, this profile could describe either a barrier 
($+ V_0$) or a trap ($- V_0$).
\medskip 

For numerical computations, using the Hamiltonian in Eq.\,\eqref{generalHam} we obtain a pair of coupled scattering-state equations within the barrier region, leading to 

\begin{eqnarray}
\label{fd2}
&& \frac{\pr}{\pr x} \, \phi_B(x) + k_y\,\phi_B(x)=\frac{i}{\hbar v_F} \,\left[\varepsilon_0(k)-V_B(x)+V_d\,\delta(x-x_s)-\Delta_G\right]\,
\phi_A(x)\ , \\
\label{fd1}
&& \frac{\pr}{\pr x} \, \phi_A(x) - k_y\,\phi_A(x)=\frac{i}{\hbar v_F} \,\left[\varepsilon_0(k)-V_B(x)+V_d\,\delta(x-x_s)+\Delta_G\right]
\, \phi_B(x)\ .  
\end{eqnarray} 
In Eqs.\,\eqref{fd2} and \eqref{fd1}, when we consider a tilted barrier under an applied electric field ${\cal E}_{0}$, we should replace $V_B(x)$ by $V_B(x)-e{\cal E}_{0}x$ within the barrier region, where ${\cal E}_{0}$ can be either positive or negative. Additionally, $k_y$ remains conserved during a tunneling process  for all regions considered, i.e., on the left of the barrier $(1)$, $x<0$, inside $(2)$, $0 < x < W_B$, and to right of the barrier $(3)$, $x>W_B$. We also introduce in Eqs.\,\eqref{fd2} and \eqref{fd1} a single scatterer of strength $ \pm V_d$ at the position $x=x_s$ within the barrier region for $0 < x_s < W_B$. The scattering problem  is such that there could be both transmitted and reflected waves with the amplitudes $t_i(\varepsilon_0)$ and $r_i(\varepsilon_0)$, $i=1,2$, in regions $(1)$ and $(2)$ but only the transmitted wave $t_3(\varepsilon_0)$ in region $(3)$. 

\medskip
\par

From a physical point of view, it is very instructive to note that under the replacements $k_y\to -k_y$ and $\Delta_G\to -\Delta_G$ for $\phi_A(x)\to\phi_B(x)$ as well, Eq.\,\eqref{fd1} transforms to Eq.\,\eqref{fd2} and vice versa. This implies a hidden chirality-time-reversal (CTR) symmetry for electron tunneling in this gapped-graphene system if its scatterer-embedded potential $V_B(x)-V_d\,\delta(x-x_s)$ remains unchanged under the barrier transformation with respect to $x\to W_B-x$ for fixed barrier width $W_B$, i.e., $V_B(x)-V_d\,\delta(x-x_s)$ should acquire a mirror symmetry with respect to its midpoint $x=W_B/2$ under zero bias ${\cal E}_0=0$ condition. 

\section{Results and Discussions}

In order to ensure that our numerically calculated results based on the finite-difference approach (FDA) are accurate and valid, we first compare them with some previously known cases having analytical solutions\,\cite{setare2010klein} for a square barrier with $V_B(x) = V_0\,\Theta(x)\,\Theta(W_B-x)$. As a validation, we have presented in Fig.\,\,\ref{f1} the transmission coefficient $T(\varepsilon,\phi_{\bf k}\,\vert\,{\cal E}_{0})$ for a square barrier using both our FDA results and the known analytical solutions\,\cite{setare2010klein}. This direct comparison clearly indicates that the FDA will be valid for an arbitrary biased barrier potential profile $V_B(x)$ on the order of $10^2$-$10^3\,$meV, including both Gaussian and triangular potential barriers embedded with a single scatterer.

\medskip
\par

\begin{figure}
\centering
\includegraphics[width=0.7\textwidth]{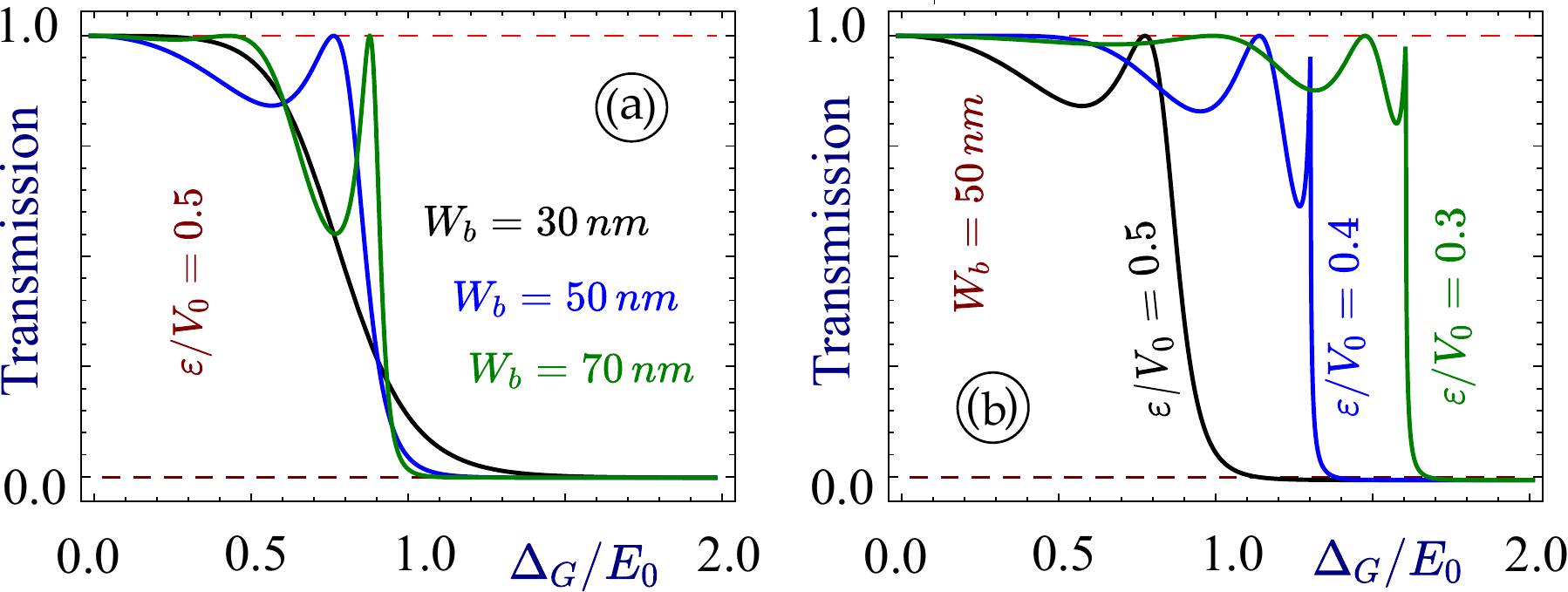}
\caption{(color online) Calculated transmission coefficients $T(\varepsilon, 0 \, \vert \, W_B)$ for the head-on collision $\phi_{\bf k} = 0$ through a square-barrier potential $V_s(x)$ as functions of bandgap parameter $\Delta_G$ with various barrier widths $W_B$ in panel $(a)$ and with different incident-electron energies $\varepsilon$ in panel $(b)$. Here, $V_0=0.2\,$eV and the unit-energy $E_0=0.1\,$eV are chosen for both panels.}
\label{f02}
\end{figure}

The analytically-calculated transmission coefficients $T(\varepsilon,\phi_{\bf k}=0\,\vert\,W_B)$ for a 1D square-barrier potential $V_{s}(x)$ are plotted in Fig.\,\ref{f02} as functions of bandgap parameter $\Delta_G$ for various barrier widths $W_B$ and incoming particle energies $\varepsilon$. We see clearly from Fig.\,\ref{f02} that the transmission coefficient $T(\varepsilon,\phi_{\bf k}=0\,\vert\,W_B)$ for head-on collision will be completely suppressed once $\Delta_G$ exceeds an $\varepsilon$-dependent threshold value, which becomes largely independent of the barrier width $W_B$. However, this threshold value for $\Delta_G$ reduces with increasing incident-electron energy $\varepsilon$.

\medskip 
\par

We now turn our attention to numerical computations of the transmission probability $T(\varepsilon,\phi_{\bf k}\,\vert\,{\cal E}_{0})$ by introducing general FDA method, as described in Eqs.\,\eqref{fd2} and \eqref{fd1}, for arbitrary shape of barrier potential profiles under an applied bias field and with embedded point scatterers.

\medskip
\par

We first present results for the effect of a barrier potential profile on the transmission probability $T(\varepsilon,\phi_{\bf k})$ in the absence of a bias field ${\cal E}_0=0$. We have specifically selected square, triangular and Gaussian as three distinctive barrier profiles in Fig.\,\ref{f2} in order to acquire a full comparison among them. From the left panel of Fig.\,\ref{f2}, we find a great suppression of Klein tunneling in the presence of a finite gap by a triangular-barrier potential $V_t(x)$ (red curve). Meanwhile, the transmission for this case is only limited to a very narrow angular region around $\phi_{\bf k}=0$. For a square barrier potential $V_s(x)$ (black curve), on the contrary, the transmission is distributed widely within a broad angular region bounded by $|\phi_{\bf k}|\leq\pi/3$, and meanwhile the transmission for head-on collision at $\phi_{\bf k}=0$ remains strong. The transmission for a Gaussian potential barrier $V_g(x)$ (brown curve) somewhat stands between the previous two cases with a limited angular distribution as well as a greatly enhanced strength at $\phi_{\bf k}=0$ compared to a square-barrier and triangular barrier potentials, respectively.

\medskip 
\par

\begin{figure}
\centering
\includegraphics[width=0.8\textwidth]{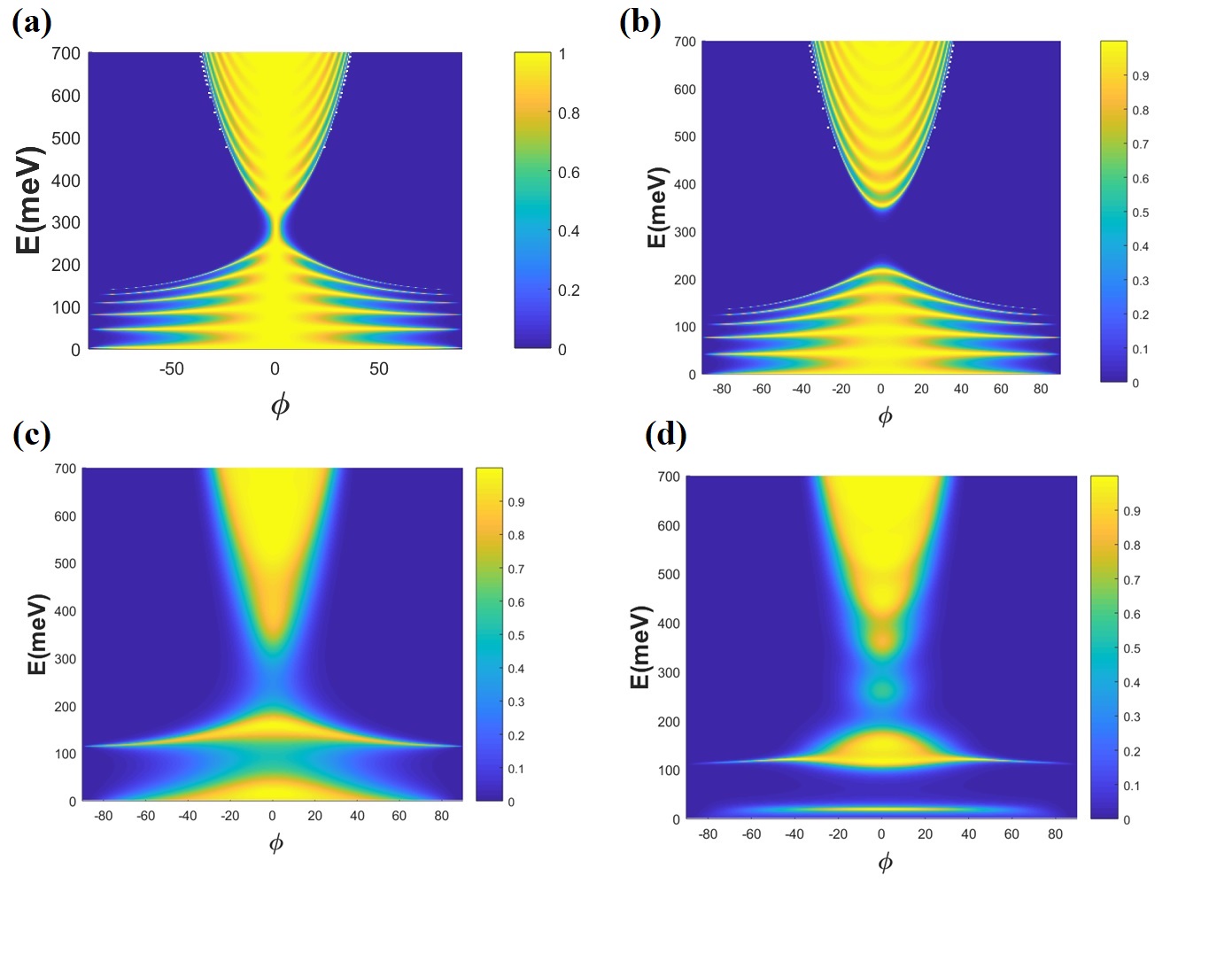}
\caption{(color online) Density plots for the transmission coefficient $T(\varepsilon,\phi_{\bf k}\,\vert\,V_B)$ as a function of the incoming-particle energy $\varepsilon$ and incidence angle $\phi_{\bf k}$ for various barrier profiles $V_B(x)$ in four different cases: (a) a square barrier without gap; (b) a square barrier with a gap $\Delta_G = 50\,$meV; (c) a triangular barrier with a gap $\Delta_G = 50 \,$meV; (d) Gaussian potential barrier with a gap $\Delta_G =50 \,$meV. For all plots, the	bias field ${\cal E}_0=0$, 
the barrier width is $W_B = 110\,$nm, and its height is $V_0=285\,$meV.}
\label{f3}
\end{figure}

In order to obtain a better and convincing understanding of the effect due to chosen potential barrier-profiles on the transmission probability,  we present in Fig.\,\ref{f3} the density plot for $T(\varepsilon,\phi_{\bf k}\,\vert\,V_B)$ as functions of incident-electron energy $\varepsilon$ and incident angle $\phi_{\bf k}$ for various barrier profiles.  Comparing Figs.\,\ref{f3}$(a)$ and \ref{f3}$(b)$, we identify a dominant feature in this figure for the tunneling of electrons in gapped graphene, i.e., the transmission probability can be substantially reduced and nearly goes to zero as the incident energy approaches the barrier height at normal incidence, where the incident particle acquires a small or even imaginary momentum within the barrier region. Additionally, the electron transmission is modified significantly for two different slowly-varying barrier profiles considered in Figs.\,\ref{f3}$(c)$ and \ref{f3}$(d)$. Here, many layered sharp resonant features of the transmission probability observed in Fig.\,\ref{f3}$(b)$ for an under square-barrier incidence disappear in both  Figs.\,\ref{f3}$(c)$ and \ref{f3}$(d)$, leaving only a single energy range around $\phi_{\bf k}=0$ in Fig.\,\ref{f3}$(d)$. 

\medskip 
\par

\begin{figure}
\centering
\includegraphics[width=0.9\textwidth]{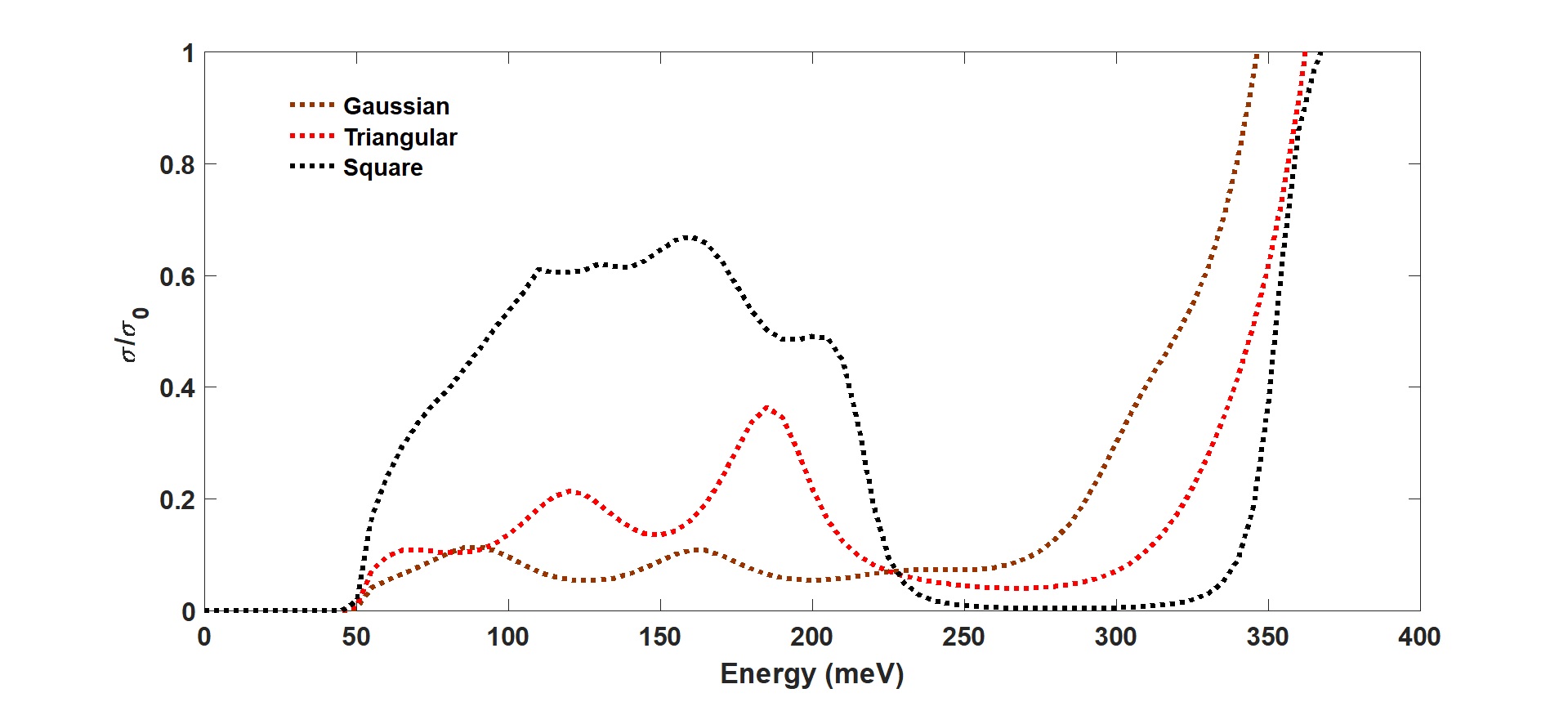}
\caption{(color online) Relative conductance $\sigma/\sigma_0$ as a function of the incident-electron energy $\varepsilon$ for three different barrier profiles, i.e., square (black), triangular (red) and Gaussian (brown) as depicted, 
in gapped graphene. Here, we set the barrier height $V_0=285\,$meV, the barrier width $W_B=100\,$nm, the gap parameter $\Delta_G=50\,$meV, and the bias field ${\cal E}_0=0$, similar to those used in Fig.\,\ref{f2}. The unit for the conductance is chosen to be $\sigma_0=2\,e^2/\pi\hbar$. 
}
\label{f4}
\end{figure}

From a technology perspective, we know that the control of an electrical current flow in graphene devices becomes crucial for their applications, such as current modulation, amplification and signal processing. For this reason, we compare in  Fig.\,\ref{f4} the changes of scaled tunneling conductance $\sigma(\varepsilon\,\vert\,V_B)/\sigma_0$ as functions of incoming-particle energy $\varepsilon$ in gapped graphene for three distinct barrier profiles. As shown in Fig.\,\ref{f4}, for $\varepsilon$ within the range of $50$-$230\,$meV, a square barrier gives rise to a square-like highest conductivity (black curve) which, however,  is decreased as $\varepsilon$ approaches $V_0-\Delta_G=235\,$meV. For a triangular barrier (red curve). On the other hand, we find a strongly-oscillating conductance with multiple peaks and valleys in the same range. Quite differently, Gaussian barrier (brown curve) leads to the lowest weakly-oscillating conductivity for all barrier profiles considered within this energy range, but it produces the highest step-rising conductivity above this energy range. Meanwhile, unlike the square barrier, the conductivity associated with either triangular or Gaussian potential barrier rises quickly for $\varepsilon>235\,$meV although its increase is not as rapid as that for the square potential barrier around $\varepsilon=350\,$meV.

\medskip
\par

\begin{figure}
\centering
\includegraphics[width=0.8\textwidth]{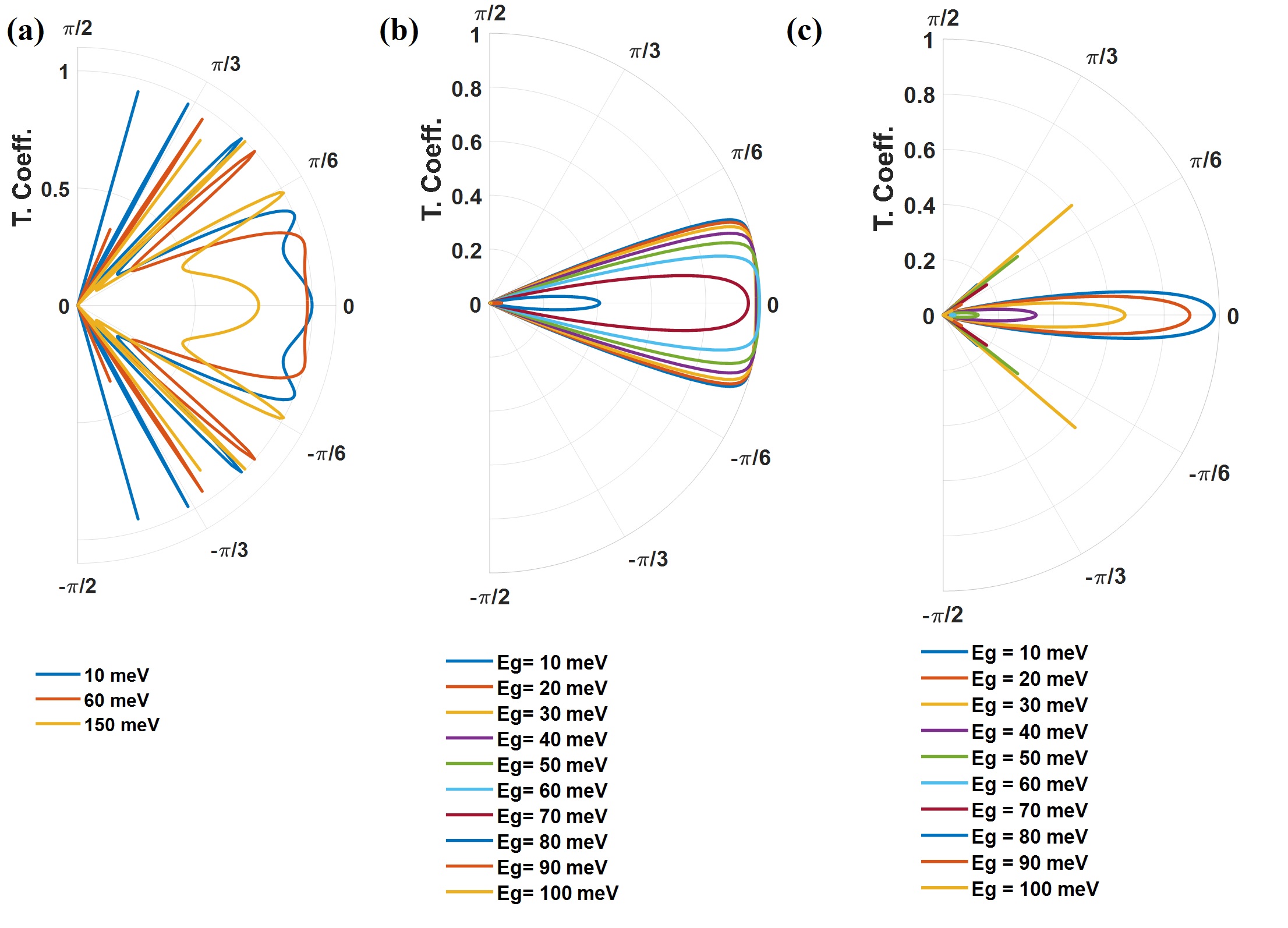}
\caption{(color online) Transmission coefficient $T(\varepsilon, \phi_{\bf k}\,\vert\, V_B)$, calculated by FDA method, as a function of the angle of incidence	$\phi_{\bf k}$ for various bandgaps in graphene and different barrier profiles $V_B(x)$: $(a)$ square barrier; $(b)$ Gaussian barrier; $(c)$ triangular barrier. Here, we set the barrier width $W_B=110\,$nm, the barrier height $V_0=285\,$meV, the bias field ${\cal E}_0=0$, and the incident-electron energy $\varepsilon=80\,$meV. 
}
\label{f5}
\end{figure}

In Fig.\,\ref{f5}, for three chosen barrier-potential profiles, we compare the obtained numerical results for the transmission probability $T(\varepsilon, \phi_{\bf k}\,\vert\, V_B)$ as a function of the angle of incidence $\phi_{\bf k}$ with a series of gap parameters $\Delta_G$. For the Gaussian barrier profile in Fig.\,\ref{f5}$(b)$, the tunneling is confined well within a small angle region. With increasing $\Delta_G$, the tunneling amplitude at $\phi_{\bf k}=0$ is enhanced quickly to unity which is accompanied by the expanded angle region around $\phi_{\bf k}=0$. Interestingly, very strong focusing of tunneling with respect to $\phi_{\bf k}=0$ is developed for a triangular barrier profile in Fig.\,\ref{f5}$(c)$ which is supplemented by the appearance of two symmetrical sharp side features. Additionally, in this case the tunneling amplitude at $\phi_{\bf k}=0$ is rapidly  reduced with increasing $\Delta_G$, which further goes together with the suppression of the two side features. For the square barrier profile in Fig.\,\ref{f5}$(a)$, lots of side peaks occur and their angle distributions are shrunken slowly as $\Delta_G$ is increased.

\medskip
\par

\begin{figure}
\centering
\includegraphics[width=1\textwidth]{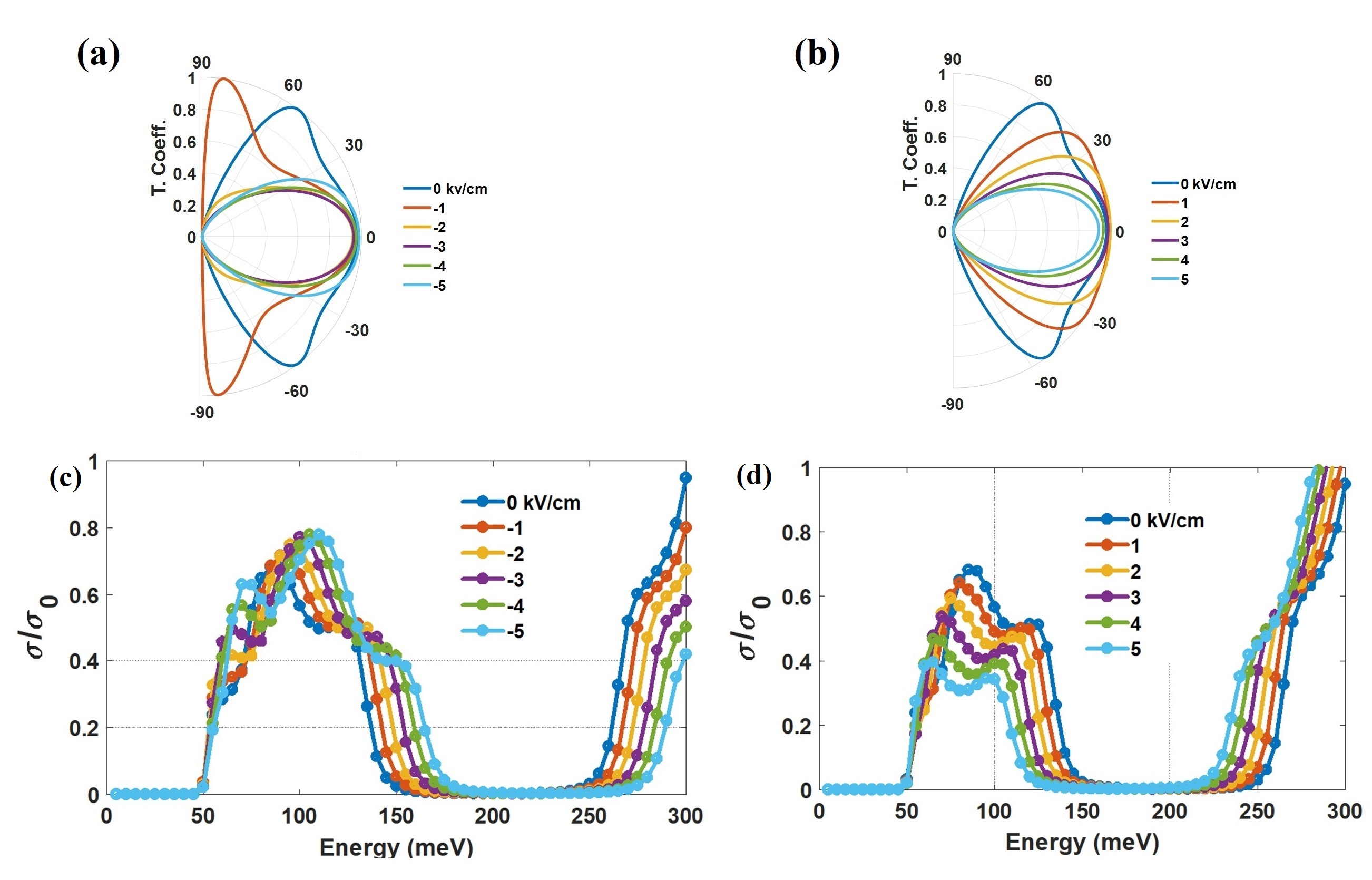}
\caption{(color online) Polar plots of $T(\varepsilon, \phi_{\bf k} \, \vert \,{\cal E}_{0})$ as a function of $\phi_{\bf k}$ for gapped graphene with $\Delta_G=50\,$meV over a square barrier with its height $V_0=285\,$meV, width $W_B=110\,$nm, as well as various applied bias fields ${\cal E}_0$. Here, Panels $(a)$-$(b)$ show the results for fixed $\varepsilon=80\,$meV corresponding to a reverse and a forward bias, respectively. Panels $(c)$-$(d)$ display the change of relative conductance $\sigma/\sigma_0$ as functions of $\varepsilon$ for a reverse and forward bias, separately. The unit for the conductance in panels $(c)$ and $(d)$ is $\sigma_0=2 e^2/\pi\hbar$. 
}
\label{f6}
\end{figure}

From a device point of view, tuning the tunneling conductance in gapped graphene is an effective technique and very important for its functionality and application. Here, we demonstrate tunable tunneling conductance for gapped graphene by applying a bias field ${\cal E}_0$ to a general barrier potential $V_B(x)$, which is in interplay with the opened bandgap of graphene. We start with a square barrier and present calculated polar plots of $T(\varepsilon, \phi_{\bf k}\,\vert\,{\cal E}_0)$ in Fig.\,\ref{f6}$(a)$-$(b)$ for forward and both backward biases  from which we find the suppression of Klein tunneling at $\phi_{\bf k}=0$ due to electron-hole transition resulting from  a finite $\Delta_G$ chosen but it is still robust against the applied bias field ${\cal E}_0$. It is interesting to note that the suppression of transmission under normal incidence appears only for a positive biase but not for a negative bias, which exhibits a strong asymmetry with respect to the polarity of ${\cal E}_0$ or broken CTR symmetry in the system. In particular, enhancements of $T(\varepsilon, \phi_{\bf k}\,\vert\,{\cal E}_0)$ near $\phi_{\bf k}\approx\pm 90^{\rm o}$ are seen only for ${\cal E}_0=-1\,$kV/cm. On the other hand, there exists an insulating zero-conductance gap for incident electron energy $\varepsilon$ which follows from Fig.\,\ref{f6}$(c)$-$(d)$, but its two edges shift in opposite directions with the polarity of ${\cal E}_0$. In this way, we can easily switch the electron tunneling conductance between the conducting and insulating phases in gapped graphene by properly selecting the polarity and magnitude of an applied bias field for any fixed kinetic energy of incident electrons.

\medskip 
\par

\begin{figure}
\centering
\includegraphics[width=0.75\textwidth]{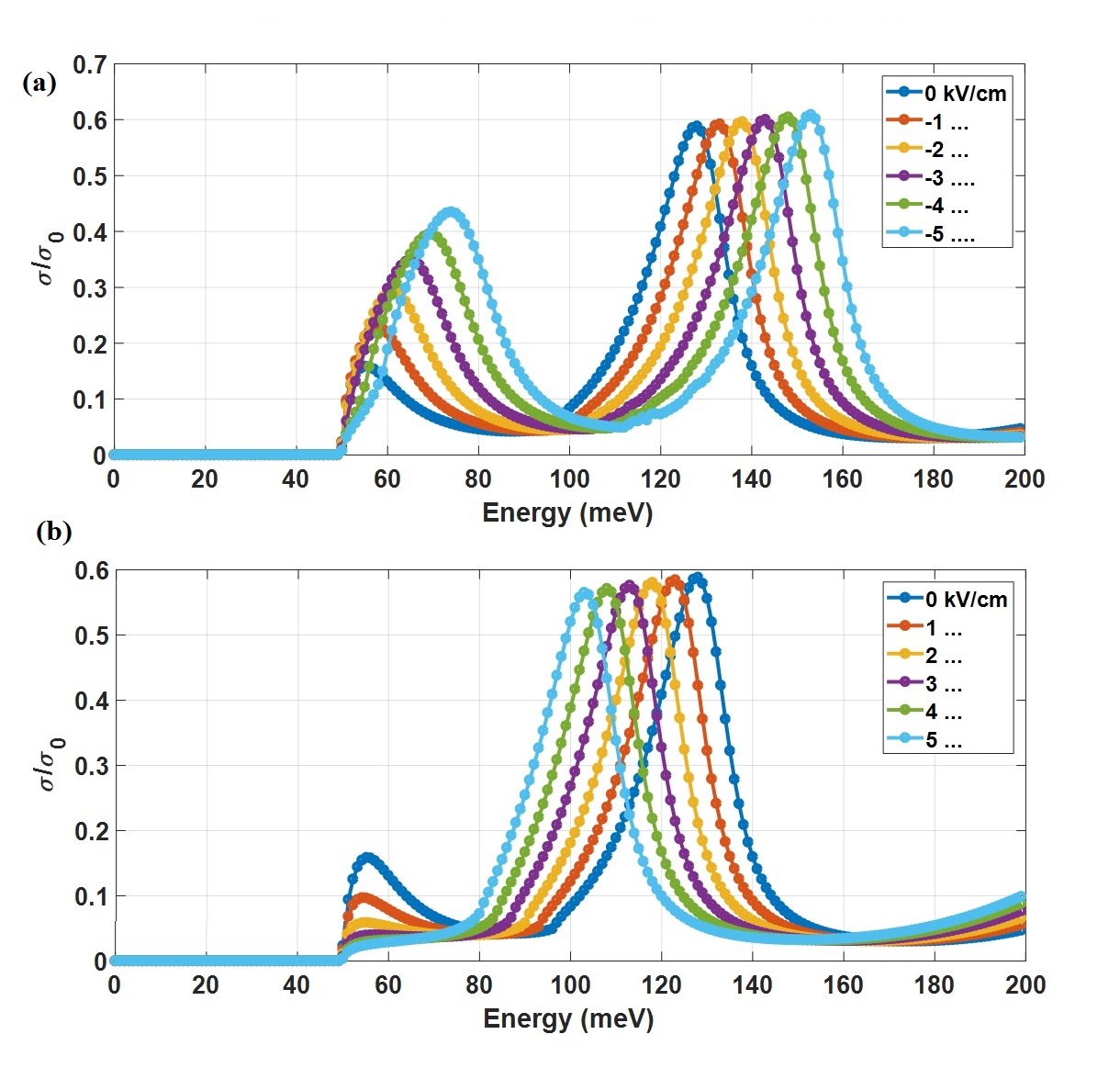}
\caption{(color online) Numerically-calculated relative conductance $\sigma/\sigma_0$  as a function of $\phi_{\bf k}$ for gapped graphene with $\Delta_G=50\,$meV over a triangular barrier with its height $V_0=285\,$meV, width $W_B=110\,$nm, as well as various applied bias fields ${\cal E}_0$. Panels $(a)$ and $(b)$ display the change of relative conductance $\sigma/\sigma_0$ as functions of $\varepsilon$ for a reverse and a forward bias, separately. The unit for the conductance in these two panels is $\sigma_0=2 e^2/\pi\hbar$. 
}
\label{f7}
\end{figure}

\begin{figure}
\centering
\includegraphics[width=1\textwidth]{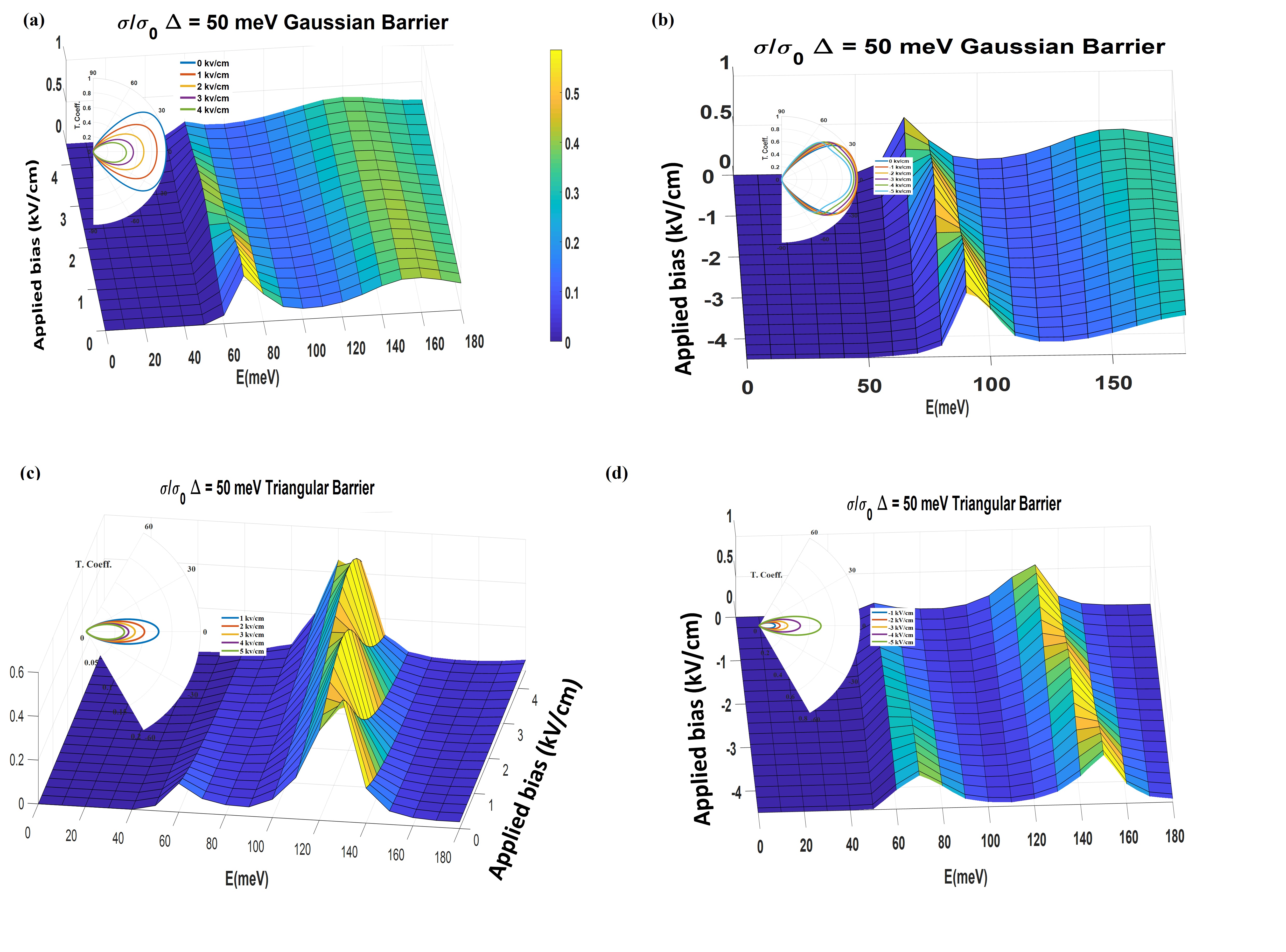}
\caption{(color online) 2D plots of tunneling conductance as functions of applied bias ${\cal E}_{0}$ and incident-electron energy $\varepsilon$. Here, four panels correspond to $(a)$ Gaussian barrier with forward bias; $(b)$ Gaussian barrier with reverse bias; $(c)$ triangular barrier with forward bias; $(d)$ triangular barrier with reverse bias. For all cases, we choose the barrier width $W_B=110\,$nm, the barrier height $V_0=285\,$meV, and the energy gap $\Delta_G=50\,$meV. The unit for the conductance in panels $(a)$-$(d)$ is set as $\sigma_0=2 e^2/\pi\hbar$.
}
\label{f8}
\end{figure}

We have established that for a finite bandgap $\Delta_G$ of graphene, the tuning of tunneling conductance by an applied bias field also depends on the selected shape of a barrier potential profile, such as a triangular or Gaussian barrier. For the triangular barrier under a reverse bias in Fig.\,\ref{f7}$(a)$, we find the conductance as a function of $\varepsilon$ remains at zero for $\varepsilon < 50\,$meV. However, both the dominant higher and the secondary lower conductance peaks move upward as a function of incoming electron energy $\varepsilon$ with increasing $|{\cal E}_0|$. On the contrary, these two conductance peaks are shifted to smaller values of $\varepsilon$ with increased ${\cal E}_0$ under forward biases, as found from Fig.\,\ref{f7}$(b)$. In order to gain an overall picture regarding the tuning of tunneling conductance $\sigma(\varepsilon,{\cal E}_0\,\vert\,V_B)$ by a bias field ${\cal E}_0$ for different barrier profiles $V_B(x)$, we compare corresponding density plots of scaled $\sigma(\varepsilon,{\cal E}_0)/\sigma_0$ as functions of both incoming-particle energy $\varepsilon$ and ${\cal E}_0$ in Figs.\,\ref{f8}$(a)$-\ref{f8}$(d)$. Generally speaking, the case for biased Gaussian barriers in Figs.\,\ref{f8}$(a)$-\ref{f8}$(b)$ only gives rise to two relatively-weak peaks in tunneling conductance $\sigma(\varepsilon,{\cal E}_0)$ within two separate ranges for kinetic energy $\varepsilon$. The bias field ${\cal E}_0$, on the other hand, only shifts those conductance peaks in $\sigma(\varepsilon,{\cal E}_0)$ downward in $\varepsilon$ under the forward-bias condition but shifts it upward as a function of $\varepsilon$ in the reverse-bias condition. For a triangular barrier, we find enhanced features for peak shifting both upward and downward with increasing $|{\cal E}_0|$, as displayed in Figs.\,\ref{f8}$(c)$-\ref{f8}$(d)$.    

\medskip
\par

\begin{figure}
\centering
\includegraphics[width=.8\textwidth]{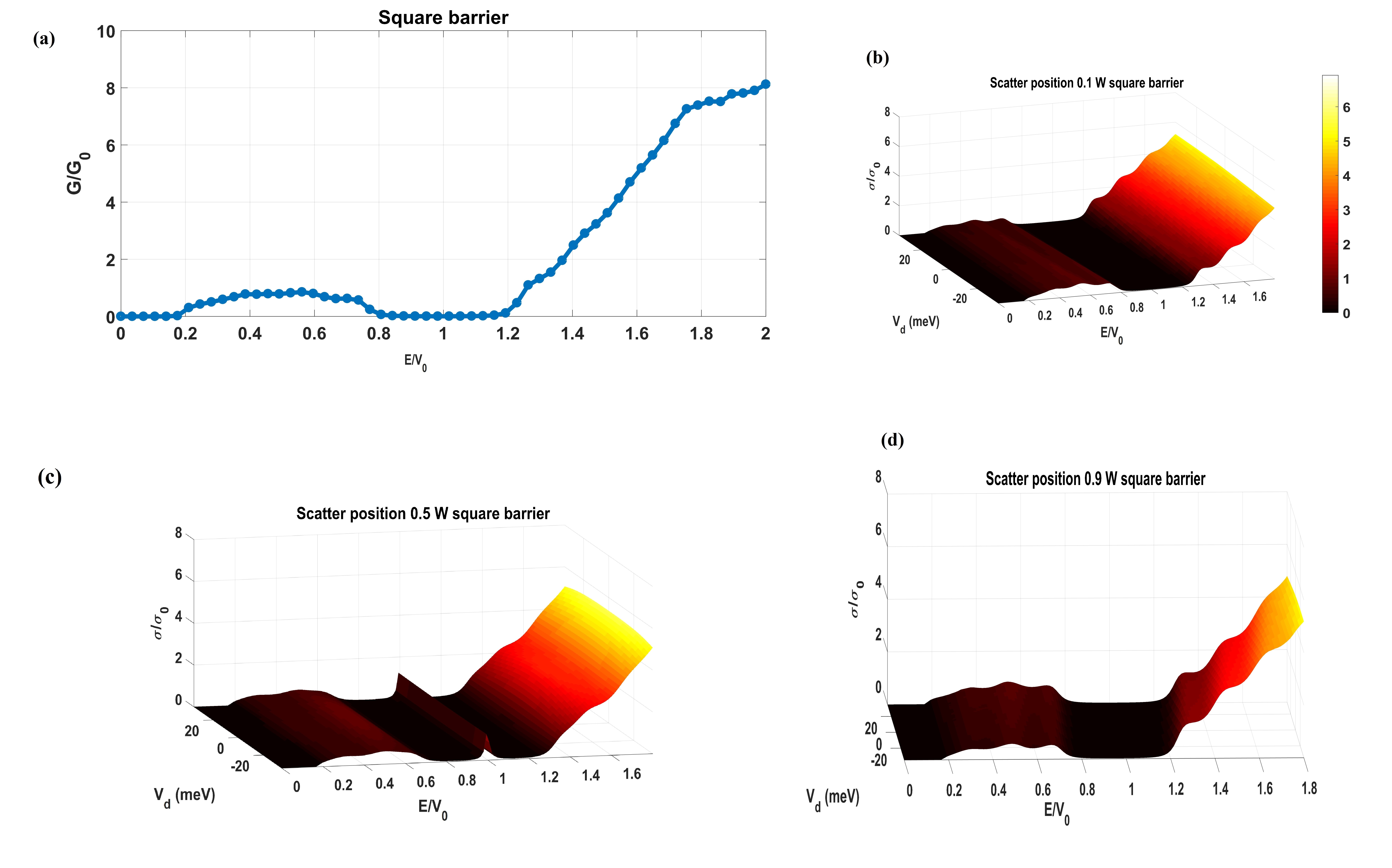}	
\caption{(color online) 2D plots for $\sigma/\sigma_0$ as functions of incident energy $\varepsilon$ and scattering strength $V_d$ (both positive and negative) for the case of a square barrier with width $W_B=100\,$nm, height $V_0=285\,$meV and zero bias ${\cal E}_0=0$. Here, panel $(a)$ presents the result for the situation with no scatterer, while panels $(b)$-$(d)$ correspond to $\sigma/\sigma_0$ with a single scatterer sitting at various positions $x_s/W_B=0.1,\,0.5,\,0.9$. The unit for the conductance in panels $(a)$-$(d)$ is set as $\sigma_0=2 e^2/\pi\hbar$.}
\label{f9}
\end{figure}

In the remaining part of this Section, we will address the effect of a single scatterer embedded at various positions within an unbiased barrier of different shapes on tunneling conductance of electrons in gapped graphene.  In Fig.\,\ref{f9}, we plot $\sigma(\varepsilon,V_d\,\vert\,V_B)/\sigma_0$ for an unbiased square barrier with different locations for a single scatterer. As a comparison, we also display the result with no scatterer in Fig.\,\ref{f9}$(a)$ under ${\cal E}_0=0$ for a square barrier potential. From Fig.\,\ref{f9}$(b)$ with $x_s/W_B= 0.1$, we observe that $\sigma(\varepsilon,V_d)$ acquires several consecutive peaks and valleys within the low-energy range of $0.3 < \varepsilon/V_0 < 0.8$ in the presence of a single scatterer. In addition, a zero-conductance gap exists for the  intermediate energy range $0.8<\varepsilon/V_0<1.3$, and meanwhile, a number of peaks and valleys develop for enhanced $\sigma(\varepsilon,V_d)$ in the high-energy region $1.3<\varepsilon/V_0<1.8$. As $x_s/W_B=0.5$ in Fig.\,\ref{f9}$(c)$ for the scatterer located right in the middle of thr square barrier, a sharp peak in $\sigma(\varepsilon,V_d)$ occurs at $\varepsilon/V_0=1$ within this zero-conductance gap, despite the sign and magnitude of $V_d$.  For $x_s/W_B=0.9$ in Fig.\,\ref{f9}$(d)$, we verify CTR symmetry with respect to the center $x_s=W_B/2$ of a one-dimensional (1D) barrier, i.e., $\sigma(\varepsilon,V_d)$ for a scatterer at $x_s$ is the same as $\sigma(\varepsilon,V_d)$ for a scatterer at $W_B-x_s$.

\medskip
\par

\begin{figure}
\centering
\includegraphics[width=.8\textwidth]{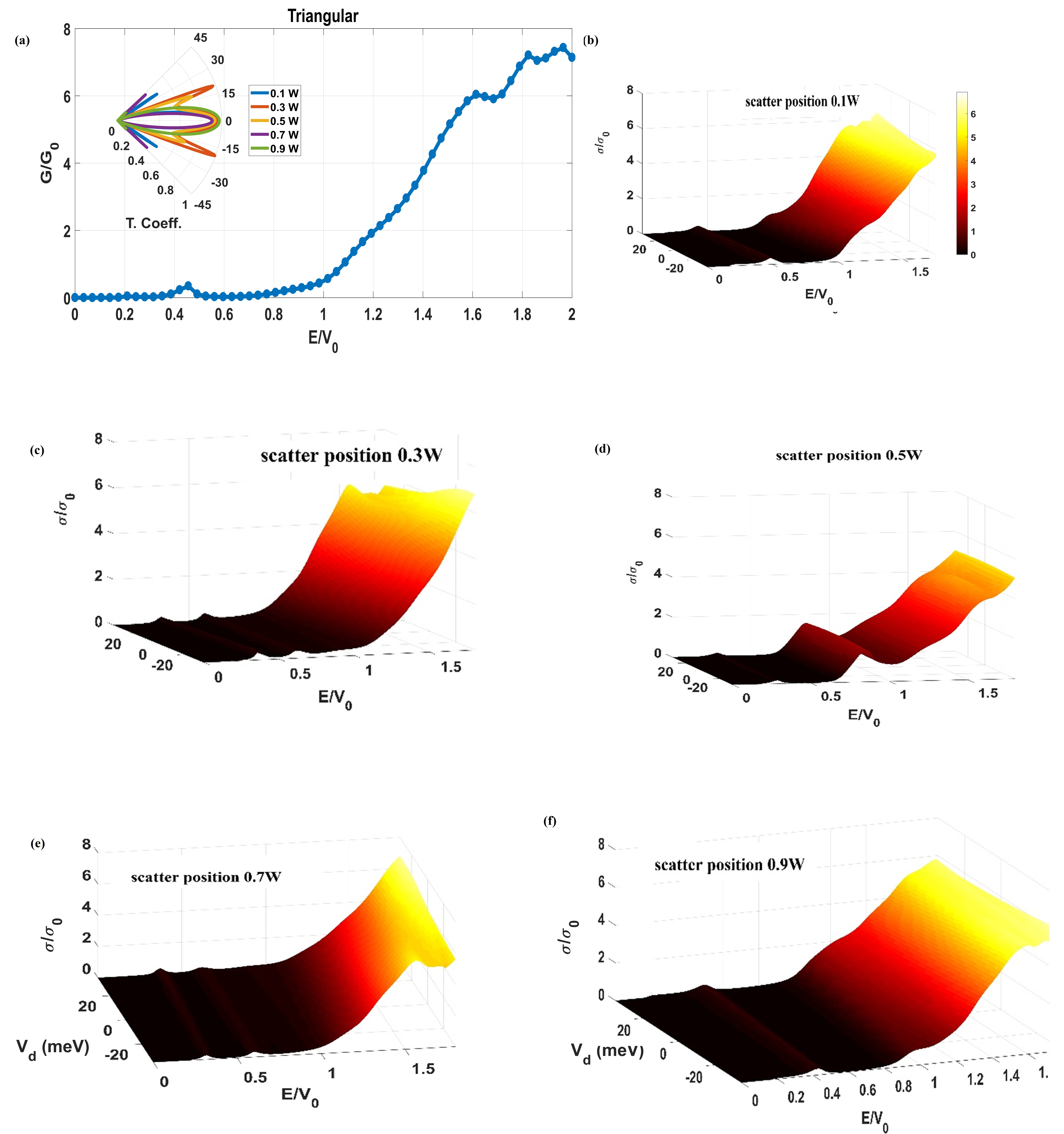}
\caption{(color online) 2D plots for $\sigma/\sigma_0$ as functions of incident energy $\varepsilon$ and scattering strength $V_d$ (both positive and negative) for the case of a triangular barrier with width $W_B=100\,$nm, height $V_0=285\,$meV and zero bias ${\cal E}_0=0$. Here, panel $(a)$ presents the result for the situation with no scatterer, while panels $(b)$-$(f)$ correspond to $\sigma/\sigma_0$ with a single scatterer sitting at various positions $x_s/W_B=0.1,\,0.3,\,0.5,\,0.7,\,0.9$, respectively. The unit for the conductance in panels $(a)$-$(f)$ is set as $\sigma_0=2 e^2/\pi\hbar$.}
\label{f10}
\end{figure}

\begin{figure}
\centering
\includegraphics[width=.8\textwidth]{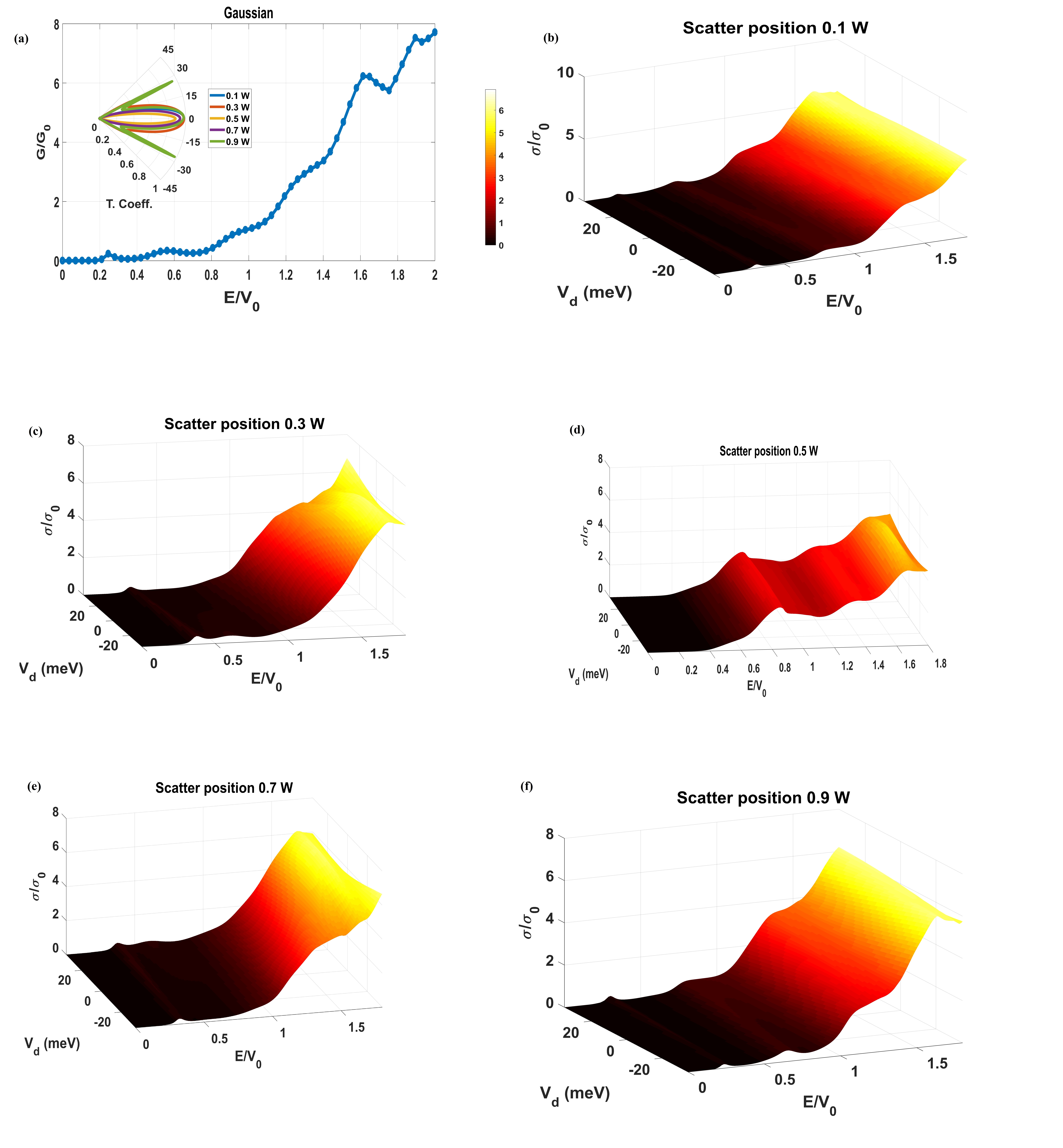}
\caption{(color online) 2D plots for $\sigma/\sigma_0$ as functions of incident energy $\varepsilon$ and scattering strength $V_d$ (both positive and negative) for the case of Gaussian barrier with width $W_B=100\,$nm, height $V_0=285\,$meV and zero bias ${\cal E}_0=0$. Here, panel $(a)$ presents the result for the situation with no scatterer, while panels $(b)$-$(f)$ correspond to $\sigma/\sigma_0$ with a single scatterer sitting at various positions $x_s/W_B=0.1,\,0.3,\,0.5,\,0.7,\,0.9$, respectively. The unit for the conductance in panels $(a)$-$(f)$ is set as $\sigma_0=2 e^2/\pi\hbar$.}
\label{f11}
\end{figure}

In order for us to understand quantitatively the interplay between effects due to a scatterer and the shape of a barrier potential, we compare 2D plots in Figs.\,\ref{f10} and \ref{f11} for the scaled tunneling conductance $\sigma(\varepsilon,V_d\,\vert\,V_B)/\sigma_0$ of gapped graphene in the presence of a single scatterer at different positions within a triangular and Gaussian barrier regions. First, for a triangular barrier in Fig.\,\ref{f10}, we observe a conductance peak at $\varepsilon/V_0=0.5$ for $x_s/W_B=0.1$ instead of $\varepsilon/V_0=1$ as in Fig.\,\ref{f9}$(b)$. Moreover, $\sigma(\varepsilon,V_d\,\vert\,V_B)$ remains zero within the energy ranges of $0 <\varepsilon/V_0 < 0.4$ as well as $0.6 <\varepsilon/V_0 < 1$. As the scatterer is shifted to $x_s/W_B= 0.3$, we find a weak conductance peak appearing near $\varepsilon/V_0=0.5$ in Fig.\,\ref{f10}$(c)$. Furthermore, when $x_s/W_B= 0.5$, we reveal a new strong conductance peak for $\varepsilon/V_0=0.85$ in Fig.\,\ref{f10}$(d)$ due to constructive superposition of two individual peaks. In particular, the CTR symmetry with respect to the center $x_s=W_B/2$ of a 1D barrier in Fig.\,\ref{f9} is still maintained in Figs.\,\ref{f10}$(e)$ and \ref{f10}$(f)$ in comparison with Figs.\,\ref{f10}$(c)$ and \ref{f10}$(b)$, respectively, with a switched sign for $V_d$.

\medskip
\par
 
Finally, for a Gaussian potential barrier, we learn from Fig.\,\ref{f11} that there exist two weak conductance peaks at $\varepsilon/V_0=0.3$ and $\varepsilon/V_0=0.65$ for $x_s/W_B= 0.1$ in Fig.\,\ref{f11}$(b)$. Additionally, zero-conductance gap is still present within the energy ranges of $0<\varepsilon/V_0<0.25$ and $0.25<\varepsilon/V_0<0.65$. Interestingly, there is a suppression of $\sigma(\varepsilon,V_d\,\vert\,V_B)$ for incident energy $\varepsilon$ below the barrier height $V_0$ as $x_s/W_B= 0.3$, as shown in Fig.\,\ref{f11}$(c)$. After $x_s/W_B$ increases to $0.5$ at the peak position of a Gaussian barrier in Fig.\,\ref{f11}$(d)$, the shape of $\sigma(\varepsilon,V_d\,\vert\,V_B)$ as a function of $\varepsilon$ changes drastically by displaying a high and wide constructive conductance peak at $\varepsilon/V_0=0.85$ accompanied by an overall enhancement of conductance in the energy range of $\varepsilon/V_0>0.4$. Furthermore, the CTR symmetry associated with $x_s=W_B/2$ for a 1D barrier is retained in Figs.\,\ref{f11}$(e)$ and \ref{f11}$(f)$ compared to Figs.\,\ref{f11}$(c)$ and \ref{f11}$(b)$ with a switched sign for $V_d$.

\section{Summary and Concluding Remarks}

In a summary, we have thoroughly investigated tunneling and calculated the crucial transmission properties of electrons across square, triangular and Gaussian potential barriers embedded with a single scatterer for gapped graphene. For this, we employed  a finite difference approach since their computations are not accessible by standard analytical solution techniques. We have also addressed the effect of a bias and point scatterer located within the barrier. It is known that the transmission and conduction in gapped graphene are largely suppressed as the particle energy lies inside the bandgap inside a barrier region. Due to the existence of a finite bandgap between valence and conduction bands, the Klein tunneling for head-on collision (i.e., with incident angle $\phi_{\bf k} =0$) is suppressed for the case with a square potential barrier. Simultaneously, the side resonances for electron tunneling, which are associated with finite incident angles, are also reduced significantly in gapped graphene. These suppression effects become even more pronounced for smooth barriers. However, we have demonstrated multiple ways to modify or even break this low-conduction condition and substantially improve the collimation of transmitted electron beam by employing the approach described above. Using the obtained transmission probability, we have further calculated the tunneling conductance which also displays a suppression for a finite gap in graphene. In fact, we have found that both the transmission and conductance display a strong dependence on the barrier profile, its slope and curvature. Meanwhile, we have also shown that the application of a bias field and its polarity greatly affect the Klein tunneling suppression resulting from broken CTR symmetry of the system, and at the same time shift conductance peaks in energy for all barrier types.

\medskip 
\par

Under a bias field, a zero-conductance gap occurs for a square barrier in a range of selected incident-electron energy below the top of the barrier. For positive/negative bias, two edges of this zero-conductance gap are respectively dragged to lower/higher energies for incident electrons with increasing absolute value $|{\cal E}_0|$ of the bias. For a triangular barrier, on the other hand,  we  have found only one dominant and another secondary peaks in higher and lower energy ranges for incident electrons. Similar behaviors of peak shifting have been seen with increasing $|{\cal E}_0|$.  By introducing a single scatterer to an energy barrier of gapped graphene, we have revealed that its strength, polarity and position can affect the conductance profile of gapped graphene. As a scatterer is moved to the midpoint (or the symmetry point) of an energy barrier, the resulting conductance always acquires either a peak within this zero-conductance gap or a significant enhancement from a constructive scattering contribution due to CTR symmetry of the system. Specifically, for a square barrier, we have observed appearance of a sharp conductance peak for $\varepsilon/V_0\sim 1$ within the zero-conductance gap. For Gaussian and triangular potential barriers, however, such a conductance peak around $\varepsilon/V_0\sim 0.85$ becomes much broadened, especially for smooth Gaussian barriers with a smaller curvature.  

\medskip 
\par

Finally,  we note that studying the conduction properties of gapped graphene and their possible alteration is very important and timely from the beginning. Practical use of graphene in device applications has already been witnessed recently. Our results are directly associated with creating spatial confinement for graphene electrons within designated areas of an electronic device modulated by a bias voltage. Apart from that, gapped graphene itself is also related to some newly discovered materials with an intrinsic spin-orbit gap, such as silicene, germanene and molybdenum disulfide. We believe that our current works could be applied to these materials as well.

\begin{acknowledgements}
D.H. would like to acknowledge the financial supports from Air Force Office of Scientific Research (AFOSR). G.G. would like to acknowledge the support from Air Force Research Laboratory (AFRL) through Contract \#FA9453-18-1-0100.            

\end{acknowledgements}

\bibliography{Farhana}

\clearpage

\end{document}